\documentclass[12pt]{book}

\usepackage[dvips]{graphicx,color}
\usepackage{makeidx,high}

\makeauthorindex
\makeindex

\begin{document}

\BookTitle{\itshape Extremely High Energy Cosmic Rays}
\CopyRight{\copyright 2002 by Universal Academy Press, Inc.}
\pagenumbering{arabic}

\chapter{Sources of sub-GZK cosmic rays}

\author{%
Peter TINYAKOV\\
{\it Institute of Theoretical Physics, University
  of Lausanne, CH-1015 Lausanne, Switzerland}\\
PeterTinyakov@cern.ch\\
Igor TKACHEV\\
{\it CERN Theory  Division, CH-1211 Geneva 23, Switzerland}\\
IgorTkachev@cern.ch}

\AuthorContents{P.\ Tinyakov and I.\ Tkachev} 
\AuthorIndex{Tkachev}{I.} 

\section*{Abstract}
We analyze the existing evidence that BL Lacertae objects (BL Lacs)
are sources of the highest-energy cosmic rays. We argue that three
independent signatures observed in the real data --- (1) improvement
of correlations with corrections of trajectories for the Galactic
magnetic field; (2) connection between $\gamma$-ray and UHECR
emissions; (3) non-uniform distribution of correlating rays over the sky ---
are consistent with the hypothesis that a substantial fraction of
cosmic rays in the energy range $(4-6)\times 10^{19}$~eV are protons
accelerated in BL Lacs.

\section{Introduction}

The fact that the observed arrival directions of UHECR [1,7,8] 
show significant clustering [7,12,9] suggests that their sources
may be identified using already existing data.  Poor (in astronomical
standards) angular resolution of the UHECR experiments and bending of
particle trajectories by Galactic and extragalactic magnetic fields
does not allow an event-by-event identification. One therefore has to
relay on statistical methods, i.g. on calculation of the angular
correlation function [9,10]. In this approach, the
significance of correlations at a given angular scale $\delta$ may be
characterized by the probability $p(\delta)$ that the observed excess
of events around sources at a scale $\delta$ may have occurred by
chance. When $p(\delta)$ is small, the correlation is physical as it
cannot be explained by a statistical fluctuation.  Apart from the
angle, $p(\delta)$ depends on the set of potential sources and other
parameters like assumed charges of particles and parameters of the
Galactic magnetic field (GMF). Significance of correlations may be
used to determine the physical values of these parameters and the subset
of actual sources by looking at the minima of $p(\delta)$.

The choice of candidate sources  depends on the assumptions
about the nature of UHECR. In the most conservative approach sources
are associated with astrophysical accelerators, among which active
galactic nuclei (AGN) may be singled out for their total energetics
and good conditions for particle acceleration [2,3]. Even in
such powerful accelerators, the acceleration of protons to energies of
order $10^{20}$~eV requires exceptional conditions. It is therefore
quite likely that only a small fraction of AGNs can produce UHECR.

Comparing the number of isolated cosmic ray events to the number of
close pairs and higher multiplicity clusters, one may estimate the
number of point sources to be of order several hundred
[4]. This agrees with the expectation that only a
special class of AGNs can be UHECR sources. A subclass of ANGs,
blazars, are thought to have relativistic jets directed along the line
of sight. Therefore, if UHECR do correlate with AGNs, they should
correlate with blazars. A subset of blazars called BL Lacertae objects
(BL Lacs) is characterized, in addition, by narrow (or absent)
emission lines. This latter property suggests low ambient matter and
radiation densities near the central engine and may provide favorable
conditions for particle acceleration to highest energies. In addition,
there are indications of negative evolution of BL Lacs (there were
less of them in the past), which makes the GZK cut-off less sharp. We
conclude that, on theoretical grounds, BL Lacs are potentially
promising sites of UHECR acceleration and should be selected as a set
of candidate sources for correlation analysis.

The most recent catalog of AGN contains 350 confirmed BL Lacs
[13], the closest being at $z\simeq 0.03$.  Our analysis
shows significant correlations between UHECR and
most powerful BL Lacs [10]. The correlations occur
at angles comparable to the experimental angular resolution. It
implies that BL Lacs are sources of UHECR.

Can one find, in the existing data, independent signatures supporting
this result? This is the main question which we address in the
present talk. We argue that these signatures do exist. Moreover, they
are compatible with the assumption that UHECR of moderate energies are
protons. 

The correlation between UHECR and BL Lacs implies that the origin of
(at least a part of) UHECR is acceleration. This has a number of
consequences. First, one should expect that protons constitute at
least part of the UHECR flux. Since protons are charged, they are
deflected in the Galactic and extragalactic magnetic fields. The
correction for these deflections should improve correlations with
sources. This effect is indeed present in the data
[11]. Second, any acceleration process is
accompanied by energy losses. A substantial part of this energy goes
into electromagnetic channel and ends up in the EGRET energy range
[2]. Thus, one may expect the connection between sources
of UHECR and gamma radiation. There are strong arguments in favor of
this connection [5].  Finally, any {\it chance
coincidences} between cosmic rays and BL Lacs should be distributed
over the sky randomly (i.e. reflecting only the local density of BL
Lacs and exposure of cosmic ray experiments).  Thus, any significant
deviation in the distribution of {\it correlating} rays over the sky
from this expectation speaks in favor of real physical connection
between cosmic rays and BL Lacs. We show that correlating rays {\it
are not} distributed randomly.

In three subsequent sections we discuss these signatures in more
detail. The last section contains discussion and conclusions.

\section{Are there protons in UHECR?}
\label{sect:GMF}

Since protons are charged, they are deflected in the Galactic (GMF)
and extragalactic (EGMF) magnetic fields. This property can be used to
detect their presence in UHECR flux if deflections in GMF dominate and
the effect of EGMF is negligible. Indeed, the Galactic magnetic field
has a regular component in which the deflections can be calculated and
actual positions of the sources (directions of rays before the
deflections) can be reconstructed. The reconstruction should improve
correlations with actual sources, and should have no effect in the
case of random coincidence.

An important ingredient in this analysis is the model of GMF. The
model we adopt consists of disk and halo components. The disk field
has the spiral structure (bi-symmetric spiral model, or BSS )
summarized in the following equations,
$$
B_\theta = B(r,\theta)\, \cos(p), \qquad
B_r = B(r,\theta)\, \sin(p),  \qquad B_z=0,
$$
where
\begin{eqnarray}
\nonumber
B(r,\theta) &=& B_0 \;\frac{R}{r \cos (\phi )}\, 
\, \cos\left(\theta - \beta \ln\left(\frac{r}{R}\right) + \phi \right),\\
\nonumber
\phi &=& \beta \ln\left(1 + \frac{d}{R}\right) - \frac{\pi}{2} \, \\
\nonumber
\beta &=& 1/\tan(p)
\end{eqnarray}
and $R$ is the distance from the Sun to the Galactic center $R =
8.5$~kpc.  The field depends on three independent parameters: local
(at the Sun position in the Galaxy) field strength $B_0$; pitch angle
$p$ which determines the direction of the local magnetic field;
distance $d$ to the magnetic field reversal.

The simplest approximation for the halo field is obtained by taking
the disk field and extending it outside of the disk with exponentially
decreasing amplitude. In addition, the halo field above and below the
disk (more precisely, its parts coplanar with the disk) may be parallel
or anti-parallel to each other. Therefore, the halo field is
\begin{equation}
B(r,\theta,z) = \sigma (z)\, \exp \left(-\frac{|z|}{h}\right)\ B(r,\theta) ,
\label{Bt}
\end{equation}
where the fourth parameter $h$ describes the extent of the halo field.
The function $\sigma (z)$ is either $\sigma(z)=1$ in the
quadrupole-type model (${\rm BSS_Q}$) or $\sigma(z)={\rm sign} (z)$ in
the dipole-type model (${\rm BSS_D}$). There are indications in
favour of the ${\rm BSS_D}$ global structure of the halo field coming
from the analysis of Faraday rotation measures of extragalactic radio
sources [6], and we adopt this model in what follows.

The values of the parameters are obtained by fitting the model to the
Faraday rotation measures of pulsars and extragalactic sources.  We
take the following values from the literature (see [11] for
details and references):
\begin{eqnarray}
&& B_0 = 1.4 \, \mu {\rm G}; \qquad p = -8^{\circ} ;\nonumber \\
&& d = -0.5 \, {\rm kpc;} \qquad h = 1.5 \, {\rm kpc}  .
\label{GMF_par}
\end{eqnarray}
Negative value of $d$ refers to the field reversal in the
direction towards the Galactic center. 

Following Ref.~[10], we take BL Lacs form the QSO
catalog [13], updated version of which contains 350
confirmed BL Lacs. The choice of UHECR set is motivated as
follows. A priori, best results should be achieved with the largest set
having best angular and energy resolution. 
Therefore, we take
all published AGASA events with energy $E>4\times 10^{19}$~eV
[7,8]. This set contains 57 events. 

The algorithm for calculation of the probability $p(\delta)$ is
described in detail in
Refs.~[9-11].  It amounts to
counting how often the excess of cosmic rays around candidate sources
observed in the real data occurs in randomly generated cosmic ray
sets.  The Monte-Carlo (MC) configurations are generated as described
in Refs.~[9-11]. The energies of MC events are taken from the
real data, while the arrival directions are generated randomly
assuming uniform distribution and taking into account the acceptance
of the experiment. In the case when particles are assumed to be
charged, each MC set is also corrected for deflections in GMF.
Exactly the same treatment of the real data and MC sets is important
to prevent appearance of artificial correlations.  Following
Ref.~[10], the angular scale $\delta$ is taken to be
$\delta = 2.5^{\circ}$. This value is used in what follows,
unless the opposite is stated explicitly.

The results of the correlation analysis are summarized in Fig.~1. Left
panel shows the dependence of the probability $p(\delta)$ on the
selected subset of BL Lacs. The selection is performed by a single cut
in apparent magnitude (objects brighter than given magnitude are
included). Three curves correspond to three charges: $Q=+1$ (solid
line), $Q=0$ (dotted line) and $Q=-1$ (dashed line). In the case
$Q=+1$, the correlations have significance of order $10^{-3}$ in a
wide range of cuts on BL Lac magnitude from ${\rm mag} < 18$ to ${\rm
mag} < 20$ which corresponds to nearly the whole set. Although jumps
in the curve suggest that fluctuations may be large, the appearance of
the ${\rm mag} = 18$ may be not a coincidence since identification of
objects may be not reliable at ${\rm mag} > 18$, see Ref.~[13]. It is also
interesting to note that at ${\rm mag} = 18$ the energy flux at optical
wavelength matches the energy flux in UHECR required of a point
source.  We adopt the cut ${\rm mag} = 18$ in what follows. At this
cut, the set of BL Lacs contains 156 objects. The left panel shows the
dependence of $p(\delta)$ on the angular scale $\delta$ in this case.
\begin{figure}[t]
  \begin{center}
    \includegraphics[height=11pc]{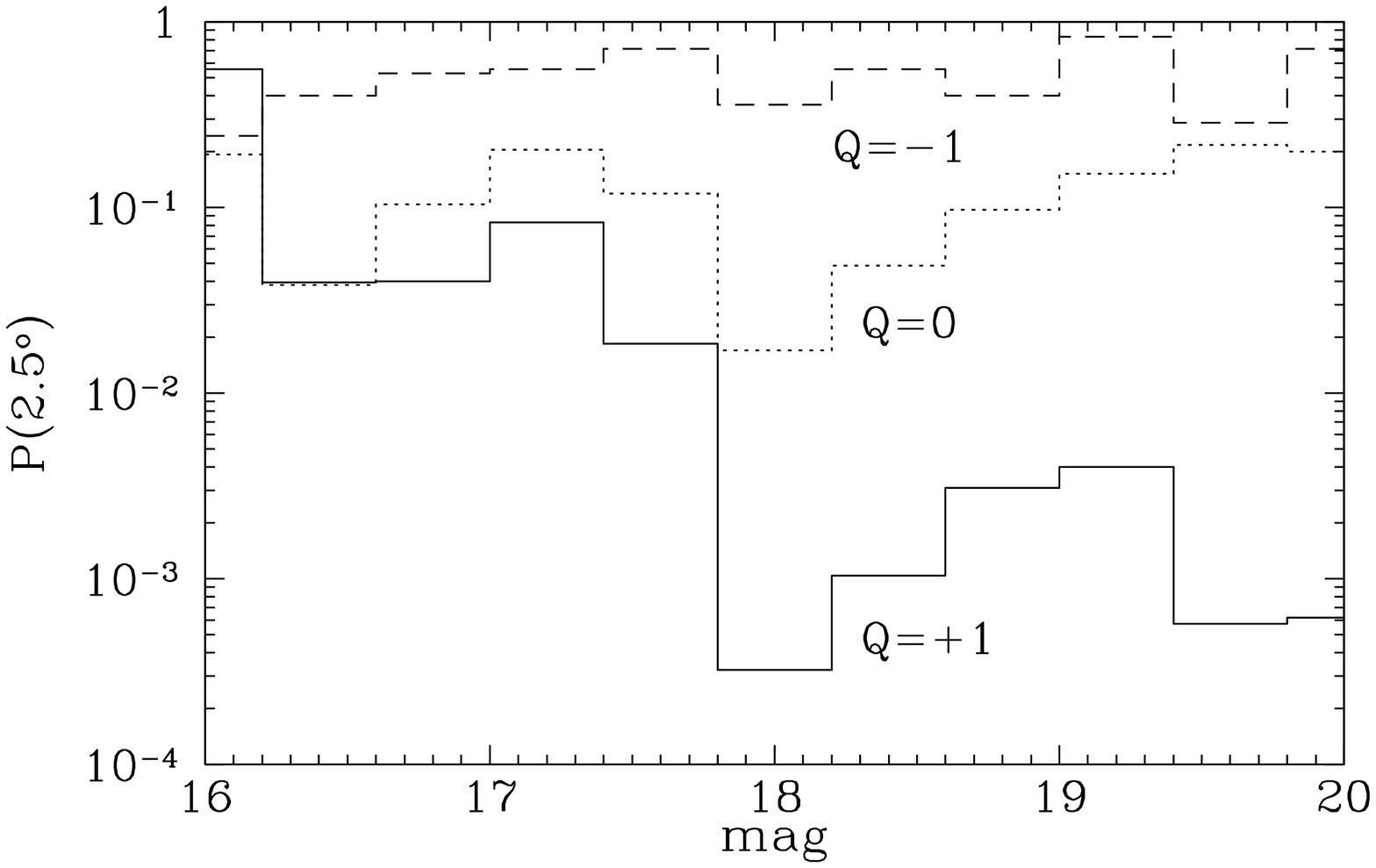}
    \includegraphics[height=11pc]{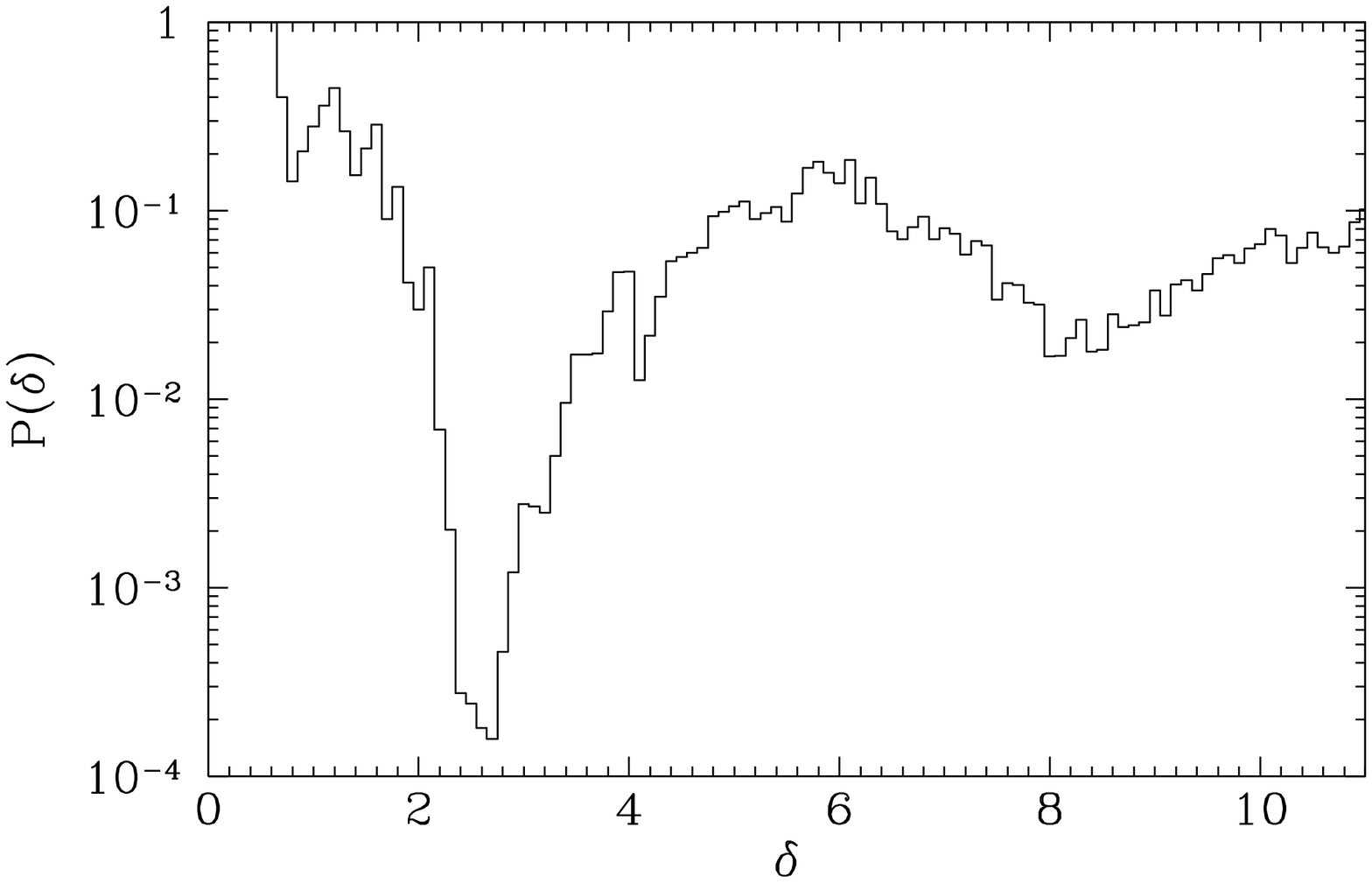}
  \end{center}
  \caption{Left panel: the dependence of the probability $p(2.5^\circ )$
    on the cut on magnitude in the BL Lac catalog. Right panel: the 
    dependence of the probability $p(\delta)$ on $\delta$ with
    the cut ${\rm mag} < 18$.}
  \label{Pdelta}
\end{figure}

\subsection{Which BL Lacs are UHECR accelerators?}
\label{sect:who}

It is likely that not all blazars accelerate particles to highest
energy, and probably not with equal efficiency. To build a realistic
model of UHECR accelerator it is very important to determine the
characteristic features of those blazars which are actual UHECR
emitters. Some hints in this direction may be found in already
existing data.

Blasars may be divided into high-polarization and low-polarization
ones. These are low-polarization BL Lacs which constitute the set
discussed above. The correlations with high-polarization blasars
(listed separately in the catalog [13]) are absent. Thus,
the low polarization may be one of the characteristic features of the
UHECR emitters.

At the minimum of $p(\delta)$ on Fig.~1 (right panel), out of total 57
cosmic rays 16 correlate with BL Lacs within $2.7^{\circ}$.  Examining
the list of BL Lacs which contribute to this correlation [11], one
notices two striking regularities.  First, the majority of objects in
this list are X-ray selected radio loud BL Lacs. Second, the fraction
of correlating BL Lacs with unknown redshifts is noticeably larger
than in the whole BL Lacs catalog. This may be explained as
follows. The absence of emission lines (more precisely, their weakness
and narrow width) is a defining feature of BL Lac family within the
general blazar class. Therefore, it is not surprising that redshifts
of roughly half of confirmed BL Lacs are not known. Increased fraction
of BL Lacs with unknown $z$ among those which correlate with the
highest energy cosmic rays may mean that the absence of emission lines
is indeed important for a blazar to be an emitter of UHECR.

Finally, one may expect the connection between UHECR and $\gamma$-ray
emissions in the sub-GeV energy region measured by EGRET. In order to
test this hypothesis, in Ref.~[5] we have selected those BL Lacs
which can be associated with gamma ray sources. The selection was
performed by looking at position coincidence within the EGRET angular
resolution. The resulting set consists of 14 potentially $\gamma$-ray
loud BL Lacs. 
Of these 14 objects 8 are found to correlate with arrival directions
of UHECR, with the significance of order $5\sigma$. This suggests that
$\gamma$-ray emission can be used as a distinctive feature of those BL
Lacs which are capable of producing UHECR.


Out of 14 associations between BL Lacs and EGRET sources, five
coincide with previously known identifications, while six are newly
proposed identifications of previously unidentified EGRET
sources. Together with the incompleteness of the BL Lac catalog, this
suggests that among the remaining {\it unidentified} EGRET sources
there may be some fraction of BL Lacs -- UHECR emitters. Their
fraction should be maximum among 90 unidentified EGRET sources which
lie outside of the Galactic plane. i.e. have $|b|>10^{\circ}$.  One
may want to add these 90 unidentified EGRET sources to the list of
candidate sources in order to increase statistics.
In what follows we show correlations of UHECR with both sets of
sources, the BL Lacs only and BL Lacs plus unidentified EGRET sources.

\begin{figure}[t]
  \begin{center}
    \includegraphics[height=13pc]{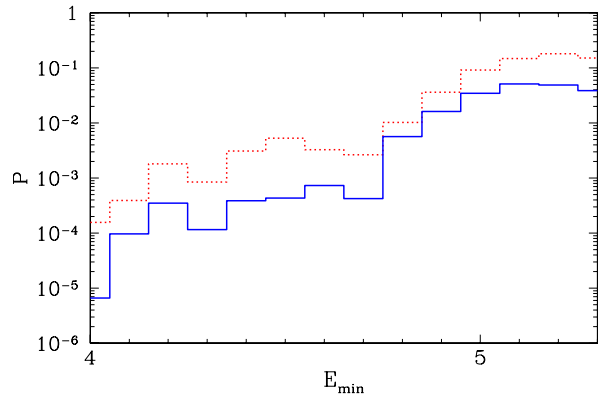}
    \end{center} \caption{The dependence of the probability
    $p(2.7^\circ )$ on the cut in energy in UHECR catalog. Dotted line
    represents correlations with BL Lacs, solid line represents
    correlations with BL Lacs plus unidentified EGRET sources.}
    \label{BLandEGRET}
\end{figure}

Fig.~\ref{BLandEGRET} shows the dependence of $p(\delta)$ on the
energy cut in the set of UHECR. The dotted curve corresponds to the
set of BL Lacs with ${\rm mag}<18$ (the point $E_{\rm cut} =4\times
10^{19}$~eV of this curve is present also on Fig.~1); the solid curve
corresponds to the same set of BL Lacs plus unidentified EGRET
sources.  We see that with the addition of unidentified EGRET sources
the correlations increase, which further supports the conjecture that
the UHECR and $\gamma$-ray emmiters are related.
We see also that correlations steadily increase with decreasing cut in
energy (i.e. with increased statistics). This implies that at $E_{\rm
min} = 4$ the extragalactic magnetic fields are yet not important for
proton propagation from BL Lacs. One may expect that this trend
persists to somewhat lower energies. 

\subsection{Global anisotropy?}
\label{sect:regions}


It is clear that {\it random} coincidences between cosmic rays and BL
Lacs should be distributed over the sky according to the acceptance of
a cosmic ray detector and local density of BL Lacs. Any significant
deviations from this distribution gives independent signature that the
correlations are real and should reflect real physical effects. There
are {\em a priori} reasons to expect such a non-uniformity. The
extragalactic magnetic fields are unlikely to be small in all
directions. If primary particles are protons, one may expect good
correlation with sources in some areas of the sky and no correlations
in the other. Poor knowledge of the Galactic magnetic field may have
similar effect: the directions of cosmic rays before they enter the
Galactic magnetic field may be obtained correctly only in the regions
where actual GMF is described well by the existing model.  Thus, one
may expect that correlating rays will not cover the acceptance region
uniformly, but will form {\it spots} where their density is high,
while in other areas the number of correlating rays will not exceed
the random background.

\begin{figure}[t]
  \begin{center}
    \includegraphics[height=17pc]{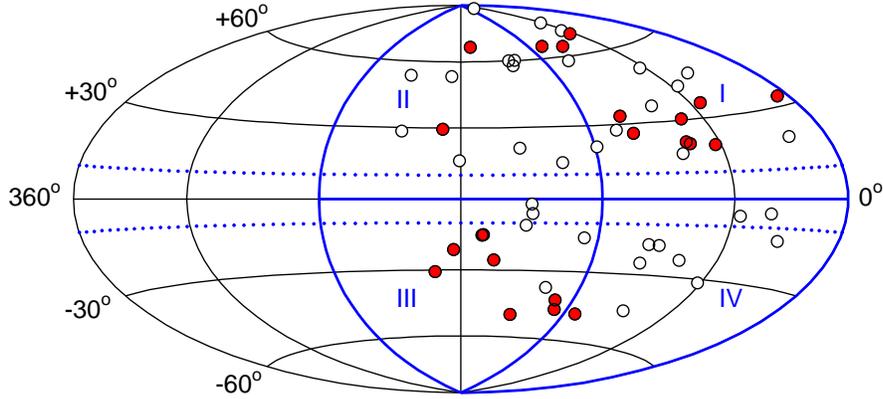}
  \end{center}
  \caption{Circles represent AGASA events with $E > 4\cdot 10^{19}$
    eV. Filled circles represent the subset of correlating rays (note
    that four of them form two tight doublets).
}
  \label{sky}
\end{figure}

To test the data against this expectation, we have divided the sky
into four regions and calculated correlation function separately in
these regions. These regions were defined in the following way.
First, we have separated Northern and Southern Galactic hemispheres
since the Southern hemisphere is sensitive to the change from $BSS_Q$
to $BSS_D$ GMF model, while the Northern is not. Second, we divided
AGASA acceptance in each of these regions in halves, one $0^\circ < l
< 120^\circ$ and the other one $120^\circ < l < 240^\circ$, in order
to separate directions to the Galactic center and to the Galactic
anti-center. We enumerate these regions from I to IV, in
counter-clockwise direction starting from the region $b>0$, $0^\circ <
l < 120^\circ$ as shown in Fig.~\ref{sky}.

\begin{figure}[t]
  \begin{center}
    \includegraphics[height=11pc]{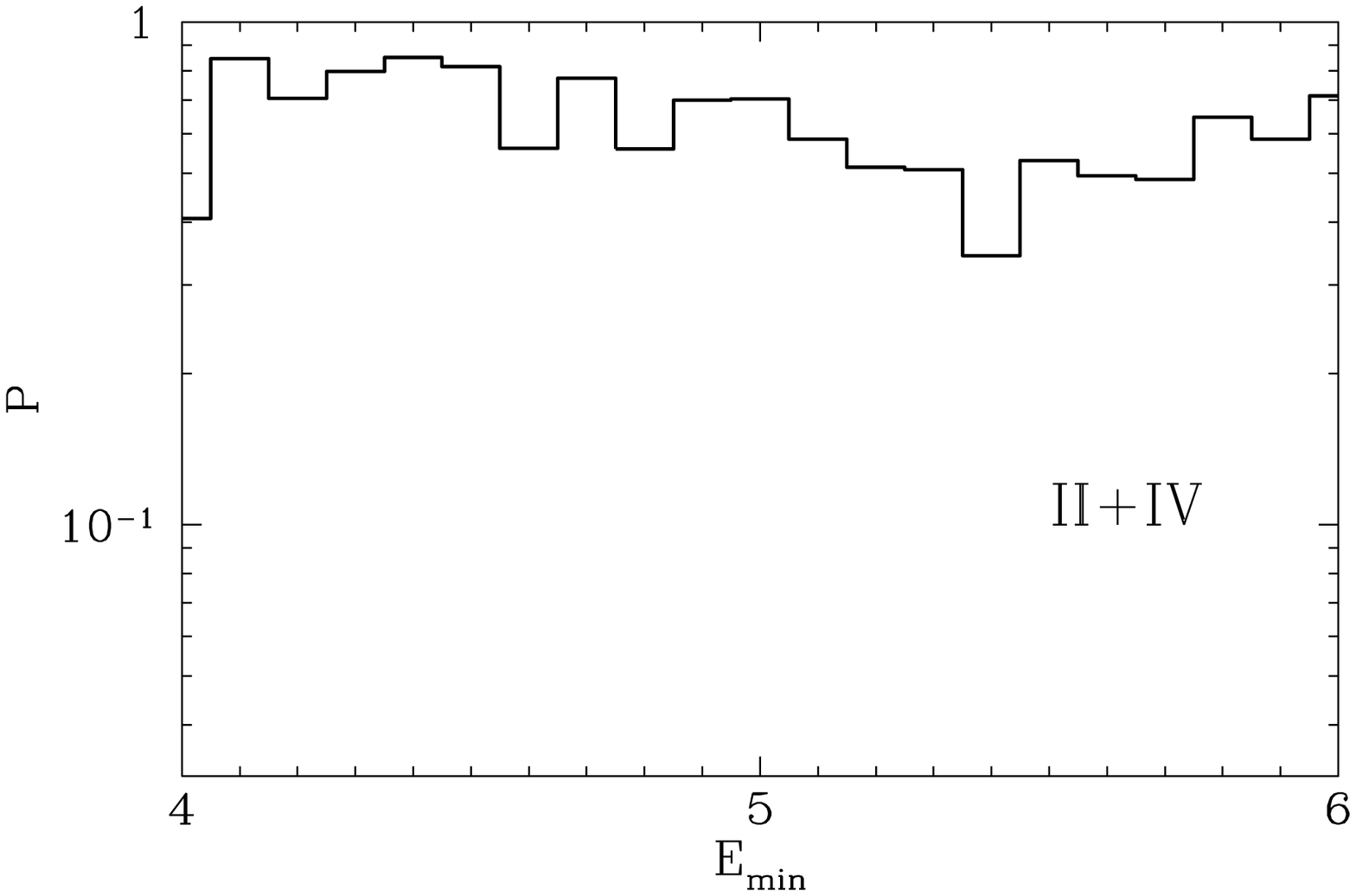}
    \includegraphics[height=11pc]{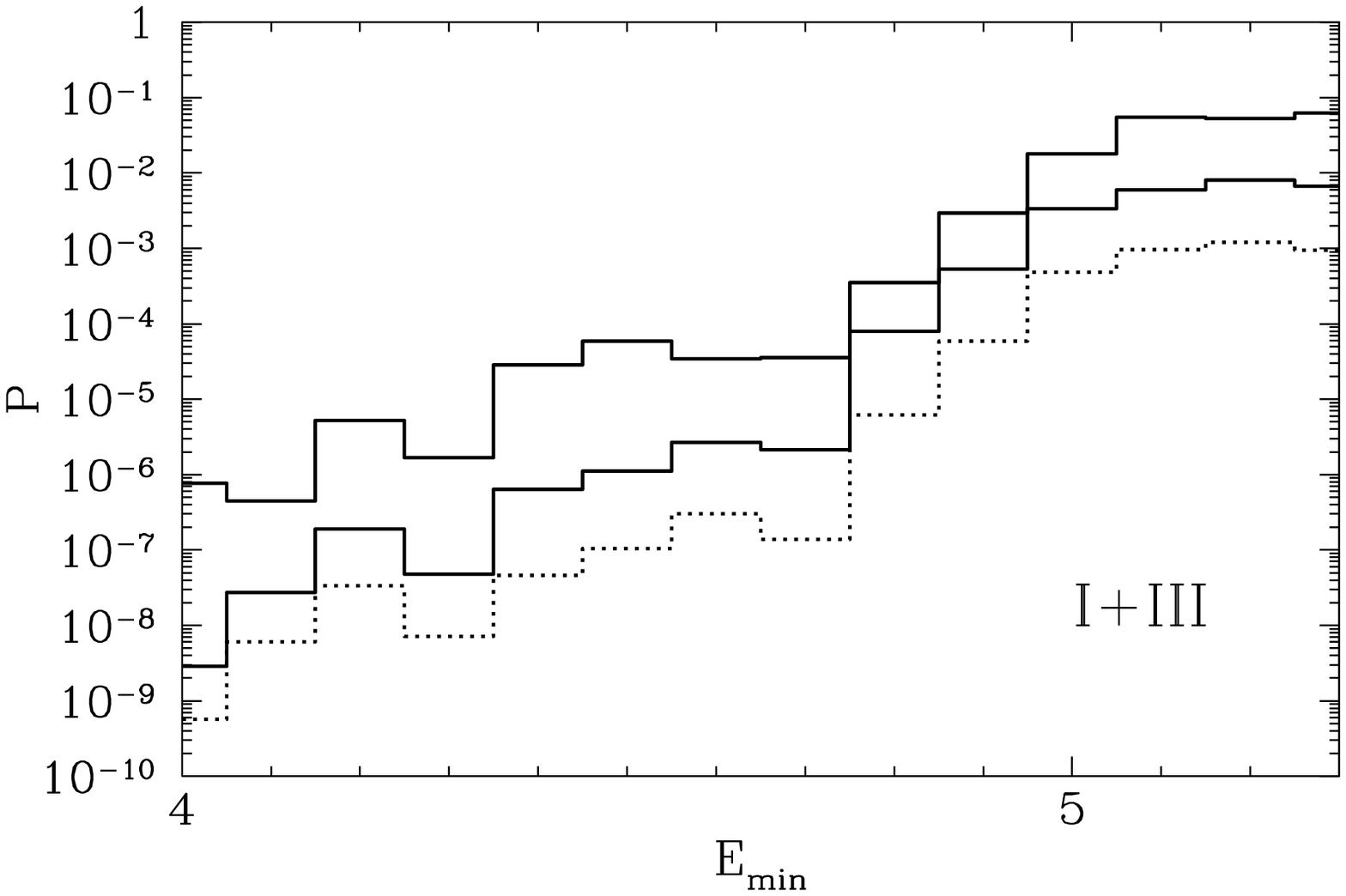}
  \end{center}
  \caption{The dependence of the probability $p(2.7^\circ )$
    on the cut in energy in the UHECR catalog. Left panel: region II+IV,
    right panel: I+III. In the right panel the curves represent (counting from
    the top): BL Lacs only, BL Lacs plus unidentified EGRET sources,
    the same but Yakutsk events are added. In any of these cases
    correlations in II+IV are absent, so only one curve is shown.}
  \label{regions}
\end{figure}

We have found that correlations in regions II and IV are absent, while
in regions I and III they are strong and have roughly equal
significance characterized by the probability of chance coincidence
below $10^{-5}$ in each of these regions separately. In Fig.~\ref{regions} we
plot correlation function calculated for the sum of regions II+IV
(left panel) and for the sum of regions I+III (right panel).

Calculation presented in Fig.~\ref{regions} takes into account the
UHECR acceptance properly. However, the observed anisotropic
distribution of correlating rays may simply reflect the anisotropy in
the BL Lacs distribution. To check that this is not the case we have
performed the following Monte-Carlo simulation. We generated cosmic
rays randomly with the distribution corresponding to the acceptance of
the experiment, and recorded those rays which correlate, by chance,
with any of the sources from the set of BL Lacs + unidentified EGRET
sources. Having accumulated a large number of such correlating rays we
counted how many of them fall into regions I, II, III and IV. Dividing
by the total number of accumulated rays we found the probabilities
that one correlating ray falls in any of these regions if the
correlation occurs by chance. For the region I+III we found
$q=0.5174$. Knowing the probability $q$ for one ray, it is
straightforward to calculate the probability of any asymmetric
distribution to occur by chance. In the case at hand, the real data
contain 22 correlating rays with 19 situated in regions I+III. The
probability to have such asymmetry by chance is $p \approx
3\cdot10^{-4}$. We conclude that the correlating rays are distributed
highly non-uniformly.

\subsection{Conclusions}

The correlations between UHECR and BL Lacs substantially increase when
arrival directions of cosmic rays are corrected for GMF. This implies
that: i)~cosmic rays of highest energies contain a substantial
fraction of protons ii)~extragalactic magnetic fields have little
effect on propagation of UHECR from BL Lacs iii)~the Galactic magnetic
field model is roughly correct at least in some directions.

The fact that UHECRs correlating with BL Lacs are not distributed over
the sky isotropically is an {\it independent} evidence of the
connection between UHECR and BL Lacs. The non-uniformity of the
distribution of correlating rays may be due to several factors: (1)
anisotropy of extragalactic magnetic fields at scales of order
$500$~Mpc; (2) poor knowledge of the Galactic magnetic field in some
areas of the sky; (3) fluctuations in the space distribution of the
closest sources. At present, it is not possible to tell which of these
factors dominate. In the future, they can be distinguished by studying
e.g. the dependence of the distribution of correlating UHECRs on
energy. However, there are hints already that at least partly this
effect is due to poor knowledge of the Galactic magnetic field.
Studying the signs of Faraday rotation measures on the map presented
in Ref. [6], we found that in regions I, III and IV the direction of
the magnetic field as given by Eqs.~(\ref{Bt})-(\ref{GMF_par}) is
correct (this does requires the use of the antisymmetric in z, dipole
type ${\rm BSS_D}$ model). However, in the region II, were
correlations are absent, the direction of the field should be
opposite.

The distribution of arrival directions suggests that the bulk of UHECR
of moderate energies are protons accelerated in BL Lacs. The numerical
simulations of UHECR energy spectrum in this model is consistent with
the data except for highest energy AGASA events [14]. If
absence of the GZK cutoff is confirmed in the future, it will have to
be explained by a separate component which starts to dominate at
energies $E> (6-10)\times 10^{19}$~eV.  At present, the correlation
analysis of arrival directions tells nothing about the nature of this
component because of low statistics.

\subsection*{Acknowledgements}

The authors are indebted to M.~Teshima for useful discussions. The
work of P.T. is supported in part by the Swiss National Science
Foundation, grant 20-67958.02.

\section{References}
\vspace{\baselineskip}


\re
1.\ Afanasiev B.~N. et al.\
1996, in Proc. Int. Symp. on Extremely High Energy Cosmic Rays:
Astrophysics and Future Observatories, Ed. by Nagano, p.32

\re
2.\ Berezinsky V.~S., Bulanov S.~V., Dogiel V.~A., Ginzburg V.~ L., 
Ptuskin V.~S. 1990, Astrophysics of Cosmic Rays (Amsterdam: Elsevier)
  
\re
3.\ Biermann P. L., Strittmatter P. A. 1987, ApJ 322, 643

\re
4.\ Dubovsky S.~L., Tinyakov P.~G., Tkachev I.~I. 2000,
Phys.~Rev.~Lett. 85, 1154 

\re
5. Gorbunov D.~S., Tinyakov P.~G., Tkachev I.~I., Troitsky S.~V.
2002, Ap.\ J.\  {577}, L93

\re
6. Han J. L.,\ Manchester R. N., Berkhuijsen E. M., Beck R. 1997,
   Astron. Astrophys. 322, 98

\re
7.\ Hayashida N. et al.
1999, Astrophys.\ J.\ {522}, 225

\re
8.\ Hayashida N. et al.,
astro-ph/0008102

\re
9.\ Tinyakov P.~G., Tkachev I.~I. 2001,
JETP Lett. 74, 1 

\re
10.\ Tinyakov P.~G., Tkachev, I.~I. 2001,
JETP Lett. 74, 445 

\re
11. Tinyakov P.~G., Tkachev, I.~I. 2002,
Astropart.\ Phys.\  {18}, 165

\re
12.\ Uchihori Y. et.\ al. 2000,
Astropart.\ Phys.  13, 151

\re
13.\ V\'eron-Cetty M.-P.,  V\'eron P. 2001,
A\&A 374, 92

\re
14.\ Berezinsky V., A.~Z.~Gazizov, S.~I.~Grigorieva
2001, astro-ph/0210095


\end{document}